%% file: sample_cref.tex
\documentclass[sigconf]{acmart}
\renewcommand\footnotetextcopyrightpermission[1]{} 
\usepackage{booktabs} 

\usepackage[caption=false,font=footnotesize]{subfig}
\usepackage{listings}
\usepackage{multirow}

\usepackage{tablefootnote}

\usepackage{graphicx}
\setcopyright{none}

\author{Crefeda Faviola Rodrigues}
\orcid{0000-0001-7401-6265}
\affiliation{%
  \institution{The University of Manchester}
}
\email{crefeda.rodrigues@postgrad.manchester.ac.uk}
\author{Graham Riley}
\affiliation{%
  \institution{The University of Manchester}
}
\email{graham.riley@manchester.ac.uk}
\author{Mikel Luj\'an}
\affiliation{%
 \institution{The University of Manchester}
}
\email{mikel.lujan@manchester.ac.uk}

\begin{document}
\title{Fine-Grained Energy and Performance Profiling framework for Deep Convolutional Neural Networks}
\begin{abstract}
There is a huge demand for on-device execution of deep learning algorithms on mobile and embedded platforms. These devices present constraints on the application due to limited resources and power. Hence, developing energy-efficient solutions to address this issue will require innovation in algorithmic design, software and hardware. Such innovation requires benchmarking and characterization of Deep Neural Networks based on performance and energy-consumption alongside accuracy. However, current benchmarks studies in existing deep learning frameworks (for example, Caffe, Tensorflow, Torch and others) are based on performance of these applications on high-end CPUs and GPUs. In this work, we introduce a benchmarking framework called "SyNERGY" to measure the energy and time of 11 representative Deep Convolutional Neural Networks on embedded platforms such as NVidia Jetson TX1. We integrate ARM's Streamline Performance Analyser with standard deep learning frameworks such as Caffe and CuDNNv5, to study the execution behaviour of current deep learning models at a fine-grained level (or specific layers) on image processing tasks. In addition, we build an initial multi-variable linear regression model to predict energy consumption of unseen neural network models based on the number of SIMD instructions executed and main memory accesses of the CPU cores of the TX1 with an average relative test error rate of 8.04 $\pm$ 5.96\%. Surprisingly, we find that it is possible to refine the model to predict the number of SIMD instructions and main memory accesses solely from the application's Multiply-Accumulate (MAC) counts, hence, eliminating the need for actual measurements. Our predicted results demonstrate 7.08 $\pm$ 6.0 \% average relative error over actual energy measurements of all 11 networks tested, except MobileNet. By including MobileNet the average relative test error increases to 17.33 $\pm$ 12.2 \%.
\end{abstract}

%
%



\maketitle

\input{main_body}

\bibliographystyle{ACM-Reference-Format}
\bibliography{sample}

\end{document}

%% file: main_body.tex
\section{Introduction}
The aim of contemporary and future computing systems is to deliver higher performance at lower power budgets \cite{parallelcomputing}. This includes embedded systems or "edge-devices" that add further limits to energy-usage due to limited battery life. Applications such as key-word spotting, facial recognition, language translation and others have become ubiquitous with the recent developments in the field of deep learning \textit{models} \cite{energysurvey}. Specifically, Convolutional Neural Networks (hereafter referred to as ConvNets) have achieved state-of-art results in various vision domains and natural language processing domains \cite{energysurvey}. 
To enable such applications for embedded devices, optimization efforts are spreads across all levels: At the \textit{algorithmic level}, newer compact neural network designs \cite{mobilenets, googlenet, squeezenet}, compression and pruning techniques \cite{compression,yangdesign}, reduced precision \cite{cour} and scheduling techniques \cite{lane2} are being proposed to save memory and increase throughput.  At the \textit{software level}, device-specific software implementations such as TensorRT \cite{tensorrt}, ARM Compute library \cite{armcompute}, Qualcomm's Snapdragon Neural Processing Engine (NPE) \cite{qualnpe}, CoreML \cite{coreml} and TensorflowLite \cite{tlite} aim to accelerate deep learning inferences or \textit{deployment} on existing mobile platforms. These libraries are complementary to existing deep learning frameworks such as Tensorflow \cite{tensorflow}, Caffe2 \cite{caffe} and others in which deep learning models must first be designed and \textit{trained}. At the \textit{hardware level}, application-specific hardware have emerged such as specialized GPUs (e.g. Jetson TX2) \cite{tx2}, FPGAs and ASICs \cite{tpu1,graphcore,wavecomputing, eie,eye,coproc}.

Despite this massive scale of efforts towards developing energy-efficient solutions to deep learning problems, there are surprisingly very few studies that measure energy for deep learning workloads \cite{cref,nvidiawhite,lane2,hotspots,analysis}. We consolidate our observations from these works and attribute the lack of adoption of energy-use as an evaluation criteria to the following reasons:
\begin{itemize}
    \item Lack of energy-measurement support in existing deep learning frameworks: Currently, popular frameworks such as Caffe, Torch, Tensorflow and others provide designers tools to benchmark their application's performance through timing measurements. There is no support for energy measurements as these are challenging to obtain consistently across platforms. They rely on the availability of power measurements facilities. Therefore, majority of the performance benchmarks (covered in Related work in Section 7), have been gathered on high-end CPUs and GPUs \cite{convnetbench, fathom,cnngpuperformance} which are not representative of platforms in the embedded domain. 
    \item Accuracy as a key metric to evaluate models: The ImageNet Large Scale Visual Recognition Challenge (ILSVRC) \cite{ImageNetChallenge} has been a major test-bed for the development of innovative ConvNet models. However, a given published accuracy is often achieved by averaging the accuracy from an ensemble of models that are executed on desktop or server systems \cite{analysis}. This implies that more computational resources are used to achieved the desired accuracy. However, on embedded platforms or specialized hardware, resource-budgets are a major concern leading to pruned versions of existing models or smaller models being chosen for deployment \cite{mobilenets,hanlearning}.
    \item Lack of systematic method and reporting standards: Measuring energy requires careful and tedious experimentation and is faced with different sources of variability. For example, power measurement facilities can vary from system-to-system. This includes different types of power meters \cite{eie,nvidiawhite,lane, caffepresso, analysis}, power sensors \cite{ti}, analytical models such as CACTI \cite{runtime,eie} and energy estimation models \cite{yangdesign}. Furthermore, various methodological choices such as  the rate at which power is sampled, baseline device power measurement, statistical validity of the measurements and measuring energy at a consistent point \cite{cref,nvidiawhite}, can lead to different results in power measurement, the details of which are often not reported.
\end{itemize}
Our work addresses the above issues by first developing a benchmark framework by integrating Caffe \cite{caffe}, a deep learning framework, and vendor-specific tools such as ARM Streamline Performance Analyzer \cite{ARMstreamline}, as shown in Figure \ref{fig:eval1}, for profiling the energy and performance (or execution time) of ConvNet models. The context for our evaluations include object recognition tasks, a single software execution framework: Caffe with back-ends like OpenBLAS \cite{openblas} for the CPU and NVidia's CuDNN \cite{cudnn} for the GPU, and a representative embedded platform such as the Jetson TX1. We do not evaluate models developed in other frameworks and focus on pre-trained models derived from Caffe's Model Zoo \cite{modelzoo}. Our methodology focuses on power measurements made using the on-board power monitoring sensor (TI-INA3221x \cite{ti} available on the Jetson TX1) for single image inferences.

Second, we use this framework for building an  initial energy-prediction model using two performance counters: SIMD instruction counts and bus accesses (or main memory accesses). The energy prediction model is built for the Convolutional (or Conv) layers in a ConvNet. We then further refine this model in an attempt to make energy predictions directly from the application parameters. To the best of our knowledge, this is the first energy estimation model available for predicting the energy consumption of all the Conv layers in a ConvNet model for the Jetson TX1. Our work shows that performance and energy are important metrics to evaluate deep learning models in conjunction with accuracy. It highlights that for the Jetson TX1, there are still unexplored ConvNet models that have high accuracy, low energy and high performance characteristics.

Finally, our work, serves as a guideline on \textit{how} to develop a systematic energy-benchmarking with the twofold objectives of 1) to understand the granularity at which power\footnote{subsequently deriving energy measures} measurements should be made on a given system, for example, system level and component level (CPU, GPU, memory) measurement, 2) understand the granularity at which power measurements should be made for the application, for example, measurement of the entire application or specific phases (in our case, layer-specific) performance and energy. 
\section{Primer on Convolutional Neural Networks}
\begin{figure}
\centering
\includegraphics[width=\columnwidth]{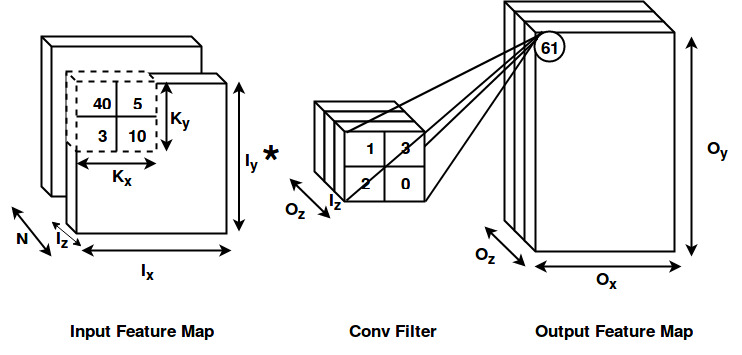}
\caption{Standard Convolution in ConvNets}
\label{fig:model}
\end{figure}

\begin{table*}[]
\centering
\caption{ConvNet models in the literature}
\label{model}
\resizebox{1.0\textwidth}{!}{%
\begin{tabular}{|l|l|l|l|l|l|l|}
\hline
\textbf{ConvNet} & \textbf{\begin{tabular}[c]{@{}l@{}}Naming\\   Convention in graphs\end{tabular}} & \textbf{\begin{tabular}[c]{@{}l@{}}Top-5\\   accuracy (\%)\end{tabular}} & \textbf{Dataset} & \textbf{\# Layers} & \textbf{Parameters} & \textbf{Model Size} \\ \hline
AlexNet & alexNet {\cite{imageNet}} & 80.3 & ImageNet & 5 Conv + 3 FC & 62 M & 244 MB \\ \hline
GoogleNet & googleNet {\cite{googlenet}} & 90.85 & ImageNet & 57 Conv + 1 FC & 6.9 M & 54 MB \\ \hline
\begin{tabular}[c]{@{}l@{}}Residual Net\end{tabular} & resNet50 {\cite{resnet}} & 93.29 & ImageNet & 53 Conv + 1 FC & 25 M & 103 MB \\ \hline
SqueezeNet & squeezeNet {\cite{squeezenet}} & 80.3 & ImageNet & 26 Conv & 1.2 M & 5 MB \\ \hline
\begin{tabular}[c]{@{}l@{}}SqueezeNet with\\  Deep Compression\end{tabular} & sqCompressed {\cite{squeezenet}} & 80.3 & ImageNet & 26 Conv & 1.2 M & 675.8 KB \\ \hline
\begin{tabular}[c]{@{}l@{}}SqueezeNet with \\ Residual Connections\end{tabular} & squeezeNetRes {\cite{squeezenet}} & 82.5 & ImageNet & 26 Conv & 1.2 M & 6.3 MB \\ \hline
VGG & vgg- small {\cite{vgg}} & 86.9 & ImageNet & 5 Conv + 3 FC & 102 M & 393 MB \\ \hline
MobileNet & mobileNet {\cite{mobilenets}} & 70.6 & ImageNet & 27 Conv & 29 M & 17 MB \\ \hline
Places-CDNS-8s & Places-CDNS-8s {\cite{placescdns}} & 86.8 & ImageNet & 8 Conv + 3 FC & 60 M & 241.6 MB \\ \hline
Inception-BN & Inception-BN {\cite{inceptbn}} & 89.0 & ImageNet & 69 Conv + 1 FC & 1.4 B & 134.6 MB \\ \hline
ALL-CNN-C & ALL-CNN-C {\cite{allcnn}} & 90.92 & CIFAR 10 & 9 Conv & 1.3 M & 5.5 MB \\ \hline
\end{tabular}
}
\end{table*}
To provide computers with ability to perform intelligent tasks such as understanding images and audio, learning, and others, the field of \textit{Machine Learning} focuses on developing mathematical \textit{models} or algorithms to acquire knowledge by extracting information from raw data. A  \textit{Convolutional Neural Network} (ConvNet) extracts information or features such as edges, color blobs through a process called \textit{feature extraction} from images, and uses this information to provide a \textit{classification} output (or a decision). It is composed of \textit{layers} to transform the raw input data into a meaningful probabilistic output. Figure \ref{fig:model}, shows the data dimensions involved in a convolution (or Conv) operation. Other typical layers found are the pooling (pool), batch norm (norm), Rectified linear unit (ReLU) and fully-connected (fc), that lend the model certain properties\footnote{This has not been explained for the sake of simplicity and the reader is advised to refer to \cite{lecunCNN} for more information.}. Each layer in a model has a certain number of computation, memory requirement and communication cost and each of these have implications in terms of energy-use \cite{iandola}.

Typically, during the \textit{training phase}, to ensure that the complexity of the model is kept in check, a time and space complexity analysis of each layer can be carried out \cite{deeperembed}. The former can be computed by counting the number of Multiply-accumulates (MAC) while the latter includes the cost for storing the input feature map ($ I_{x} \times I_{y} \times I_{z}$) to each layer, the corresponding filter weights ($ K_{x} \times K_{y} \times I_{z} \times O_{z} $) and biases and the output feature map ($ O_{x} \times O_{y} \times O_{z}$) dimensions. $N$ denotes the batch size. Here, x, y and z represent the Cartesian axes. The computational cost of a standard Convolution operation is given by:
\begin{equation}
\label{compeq1}
    O_{x} \times O_{y} \times O_{z} \times K_{x} \times K_{y} \times I_{z}
\end{equation}
and the storage cost or bandwidth (in bytes) is given by:
\begin{equation}
\label{compeq2}
     (I_{x} \times I_{y} \times I_{z} + K_{x} \times K_{y} \times I_{z} \times O_{z}+ O_{x} \times O_{y} \times O_{z}) \times 4 
\end{equation}
Tools to extract this from Caffe's ConvNet definition file are emerging \cite{caffe2any}.
Current ConvNet architectures, given in \autoref{model}, have parameters or weights typically in the order of a million. These weights are stored in 32-bit floating point precision, implying a model size of the given network to be four times its number of parameters. Large model sizes limits the storage of the parameters entirely on current on-chip SRAMs \cite{hanlearning}. Therefore, there is growing interest to develop compact sized models and prune existing models to fit on fast and low energy on-chip cache memories as well as reduce the number of computations performed \cite{googlenet,squeezenet,resnet,compression,mobilenets}. For example, topologies such as the \textit{fire} module in SqueezeNet, \textit{inception} module in GoogleNet and \textit{depth-wise separable convolutions} in MobileNet\footnote{Note this is a re-implementation in Caffe \url{https://github.com/chuanqi305/MobileNet-SSD}} aim to reduce the number of computation of the model by targeting the Conv layers. The computational cost of a depth-wise separable convolution is computationally more efficient than the standard convolution \cite{mobilenets} and is given by:
\begin{equation}
\label{compeq3}
    O_{x} \times O_{y} \times K_{x} \times K_{y} \times I_{z} + I_{z} \times O_{z} \times O_{x} \times O_{y} 
\end{equation}
While model sizes are reducing to bring the energy-costs down another important aspect is the managing the data movement between layers.
Table \ref{model} gives a list of the models chosen for this study where Column 5 represents the cumulative counts of two types of layers: Conv and fc present in each model. Recently, in newer ConvNet models the use of global average pooling layer has been used instead of traditional fully-connected layers \cite{nin}. In our study, we target the Conv layers as a first step towards developing an energy prediction model. Later, we envision extending the methodology to build more sophisticated predictors to predict energy for other types of layers. Our work reports coarse-grained energy consumption of models such as AlexNet, SqueezeNet and GoogleNet and compare it against values reported in the literature \cite{nvidiawhite}. We also include other models that have no known reported energy measurements, given in \autoref{model}. The top-5 accuracy of the model is the top-5 predictions of the object category in a given image from the ImageNet dataset \cite{ImageNetChallenge} and is a measure of how well the model performs for the task of image classification. ALL-CNN-C is typical model for smaller datasets like CIFAR-10 dataset.
\section{Power and performance measurements}  
\begin{figure}
\centering
\includegraphics[width=\columnwidth]{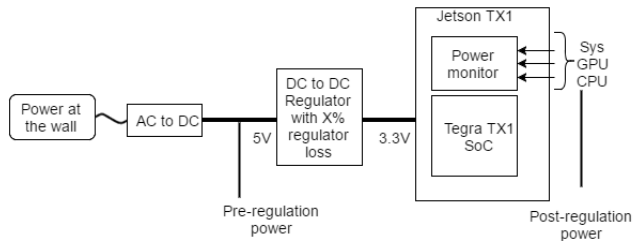}
\caption{Power monitor on the Jetson TX1}
\label{fig:power}
\end{figure}

\begin{figure}[!t]
\centering
\subfloat[GPU, CPU, System power profiles]{\includegraphics[width=\columnwidth]{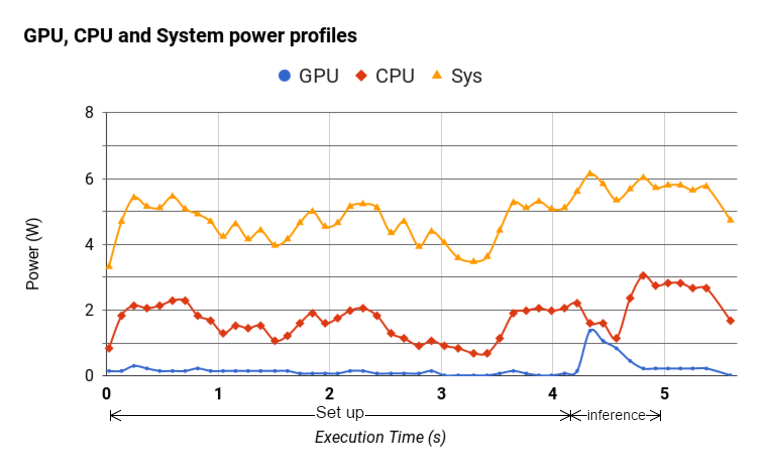}
\label{fig:prof2}}
\hfil
\subfloat[GPU power profile with GPU frequency]{\includegraphics[width=\columnwidth]{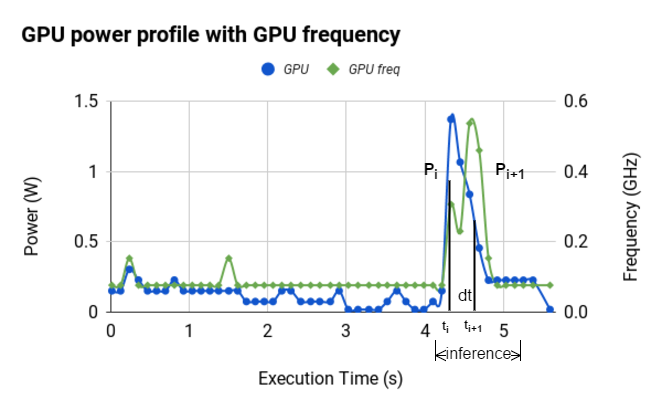}
\label{fig:prof}}
\caption{Sample power and frequency profiles with inference running on the GPU}
\label{fig:profs}
\end{figure}

The embedded system chosen for the power and performance measurements is the Jetson TX1
which has a 256 CUDA core Maxwell GPU running at 1GHz and a quad-core of ARM Cortex A57/A53 running at 1.9GHz. By default, the TX1 can be cross-compiled with the Linux kernel version (3.10.96) using Jetpack 2.3 and has a host operating system of Ubuntu 16.04. However, to enable power measurements, the kernel has to be modified to include the TI-INA3221x power monitor that is shipped with it. This power monitor provides system level power at VDD\_IN , CPU level power at VDD\_CPU and GPU level power at VDD\_GPU, as shown in Figure \ref{fig:power}. This is the post-regulation power after AC to DC conversion of the wall power and DC to DC conversion (pre-regulation power conversion) as required by the system-on-chip (SoC). Power values (mW) are instantaneous and are accessible from the hardware counters on the \textit{/sys} file system (\textit{sysfs}) which is read by our sampling script.
Given, the ability to measure power using the monitoring chip at different levels, a sample power profile of an inference executing on the mobile GPU is shown in the Figure \ref{fig:prof2}. Note, that the chip does not provide a facility to explicitly measure DRAM power.
Here, the x-axis represents execution time for the entire application and the y-axis represents the specific power values read from the different counters using a sampling rate of the sampling software or the user sampling rate. In addition, Figure \ref{fig:profs} also shows the various phases of the application such as the set up phase running on the CPU and the inference phase running on the GPU (see Section 4 on how to map application phases to the power profile). The variation in power values over the entire duration of the application is a result of the techniques that exists for power management in embedded systems including Dynamic voltage and frequency scaling (DVFS) techniques that take into account the characteristics of the application (for example if a task is memory bound frequency can be lowered) to gate cores at specific voltage or frequency levels and Power mode management (PMM) that can make use of idle time intervals in an application to switch specific components into low power modes when not in use \cite{mittal}. The Jetson TX1, allows dynamic power scaling of its CPU as well as GPU through the use of its default governors \cite{jetsongov} and this GPU frequency scaling behaviour as the inference application executes, is seen in Figure \ref{fig:prof}.

Let us consider a deep learning application running on the GPU, the energy consumption of the inference is the area under the peak of the GPU power curve in Figure \ref{fig:prof} and can be calculated by Equation \ref{eq1}:
\begin{equation} \label{eq1}
E_{inference}=\sum_{i=0}^T E_{dt}= P_{i+1} \times dt 
\end{equation}
where $E_{dt}$ is the area of the rectangular strip shown and is calculated using the $i+1^{th}$ power sample $P_{i+1}$ over that duration $dt= t_{i+1}-t_{i}$. The total energy for the inference $E_{inference}$ is the sum of all the rectangular areas over the duration of the inference $T$. In our study, we report the execution time per image (s/image) or performance and the energy per image (J/image) for single image inferences.
\section{Evaluation framework}
\begin{figure}
\centering
\includegraphics[width=\columnwidth]{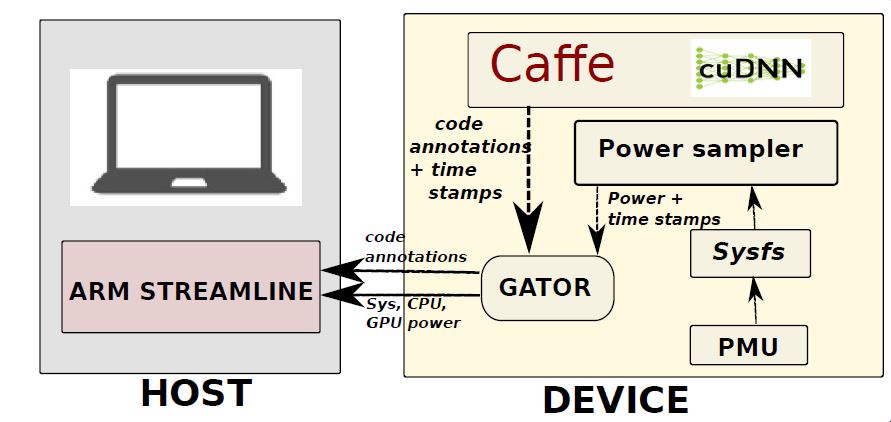}
\caption{Overview of the evaluation framework}
\label{fig:eval1}
\end{figure}
The evaluation framework is composed of two components: the hardware component that includes the target device and the power monitor that provides direct power values, and the software component as shown in Figure \ref{fig:eval1} that implements the methodology to acquire accurate and consistent power and performance measurements. 
The software component is divided into three distinct parts:
\begin{itemize}
\item Many existing deep learning frameworks such as Caffe, Tensorflow, Torch, Theano and others allow users to define the software environment within which the inference application is executed \cite{surveyefficientdnns}. It is used to define the type of inference application, the choice of algorithm to use, the implementation of the algorithm, device selection, for example, CPU or GPU and the number of inferences.
\item The power sampling method defines the procedure used to collect power samples and the experimental set up such as the user-defined sampling rate, the devices conditions when these measurements were made, for example, interaction with the device power management or DVFS system, the baseline power and idle conditions.
\item The post-processing step processes this raw power to derive the energy measurements at specific phases of the application or the device component level. 
\end{itemize}
\subsection{Deep learning framework}
To carry out an inference, we chose the Caffe framework \cite{caffe} as its C++ interface allowed integration with ARM Streamline tool. Recent work such as Fathom \cite{fathom} show that most of these frameworks share similar characteristics in terms of the underlying implemented operations. Hence, such a methodology could be adopted for other frameworks as well. We chose to execute all the ConvNet models in Caffe's master branch. Caffe has experimental support for other classes of deep learning models such as Recurrant Neural Networks and  Long Short Term Memory network (LSTMs), hence, we focus on only Caffe's ConvNet models.
Our Caffe (version 1.0.0-rc3) was compiled for the GPU with Cuda (8.0) and CuDNN (5.1.5) and CPU with OpenBLAS (libopenblas\_cortexa57p-r.0.2.20.dev.a) with a max num-thread of 4. The application selected was an inference using a ConvNet model on a single RGB image (224 x 224 pixels) taken from ImageNet dataset \cite{ImageNetChallenge}. All computations are 32-bit floating point.
Caffe's command line test interface was used to run an inference with arguments for the model \textit{prototxt} file (deploy.prototxt), pre-trained weights file (.caffemodel), device selection flag (GPU=0) for a single iteration. When selecting CPU for inferences, the number of threads was varied by setting the environment variable OMP\_NUM\_THREADS before calling Caffe's test interface. The power monitor provides the CPU power averaged power over the four cores. The pre-trained weight file for the selected ConvNet models, given in Table \ref{model}, is available in Caffe's model-zoo repository and this weight file was stored on the local disk on the Jetson TX1. 
\subsection{Power sampling method}
The power sampling method integrates vendor-specific tool such as ARM's Streamline Performance Analyser that provides CPU hardware performance counter values such as instruction per cycle, cache misses, CPU activity and others. This tool runs on a host laptop system with Ubuntu 14.04 and remotely communicates the target device. However, by default this tool is not configured to provide power sample values and requires a custom C++ function to be defined to read power values from the \textit{sysfs} interface on the target device. The target device has to be set up to communicate with the Streamline tool with the help of a gator module and gator daemon \cite{gator}. The gator module was built as a loadable kernel module in the modified Linux kernel used for the Jetson and the gator daemon was cross-compiled on the host for the TX1. ARM Streamline can be configured for Mali-based GPUs, however since the Jetson comes with an NVidia GPU, additional tools such as $nvprof$ would have to be used in conjunction with this tool to generate GPU profile traces and correlate these to energy measurements obtained. \textit{Nvprof} provides us with information related to the type of kernels running on the GPU, GPU utilization and other metrics. Our work differs from prior works that have used GPU based profiling tools such as \textit{nvprof} to analyse the performance of ConvNets \cite{cnngpuperformance} or existing performance benchmark on desktop GPUs \cite{convnetbench}, where we restrict our studies to fine-grained energy and performance measurements on the CPUs.
\subsubsection{ARM Streamline: Mapping per layer functions to power profiles}
Studies such as \cite{kimcomp}, show the usefulness in mapping the power profile of the inference to its individual layers but omit the method used for the mapping process while other studies show the benefits of exploiting per layer energy information for better pruning \cite{yangdesign} or better heterogeneous scheduling techniques \cite{deepx}. Therefore, we provide the following method to map per layer functions in Caffe by integrating ARM's Streamline Performance tool into our evaluation framework:
\begin{figure}
\centering
\includegraphics[width=\columnwidth]{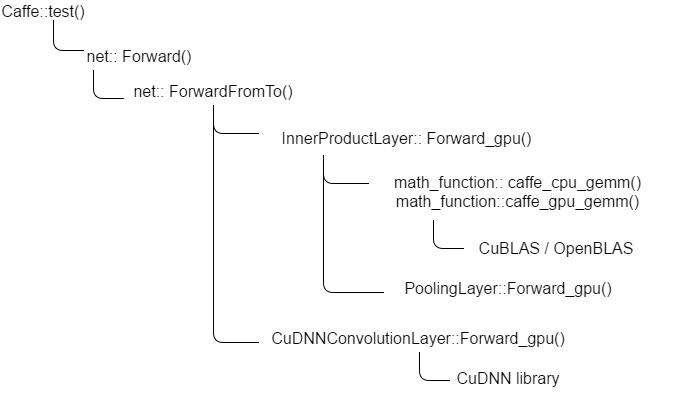}
\caption{Function structure in Caffe}
\label{fig:structure}
\end{figure}
\begin{itemize}

\item In the first step, we identify the specific functions of the algorithm, as shown in Figure \ref{fig:structure}, and map these functions to the layer-wise interpretation of a ConvNet model. This step is necessary as it helps us decide where the ``annotations or markers''\footnote{The markers can be used to denote the start and end of a function call} of ARM Streamline can be placed. The inference phase in Caffe begins by calling Caffe's $test()$ function that creates an object of type $net$ to provide the details for the model architecture and storage of the pre-trained weights. To run the actual inference it calls the $net$ class function $Forward()$. Depending on whether a loss needs to be calculated (the loss is usually calculated during the training phase) it calls $ForwardFromTo()$. At this stage there are three possible paths during the inference. Caffe can be configured to call CuDNN library for certain layers (for example, a convolution layer) if the execution runs on the GPU or Caffe's own implementation of a layer (for example, a pooling layer) or an implementation (for example, a fully-connected layer known as \textit{inner product} in Caffe)  that depends on other external back-ends of the GPU and CPU. In this work, we demonstrate  mapping of energy profiles to the highest level function call of a layer at the $net::ForwardFromTo()$ level and provide measurements for particular layer type. However, for cases such as SqueezeNet and GoogleNet that are composed of layer modules, one could insert the markers at lower levels in the hierarchy to extract measurements of its Conv or pooling layers.
\item To integrate ARM's Streamline tool with Caffe, we insert markers in the function of interest with the help of macros \textit{ANNOTATE\_SETUP} and \textit{ANNOTATE\_MARKER\_COLOR\_STR}. 
\item We create custom counters that can read the power values from $sysfs$ interface. This is done by configuring Streamline to read VDD\_IN, VDD\_GPU and VDD\_CPU on the target device, using Streamline annotation macros such as \textit{ANNOTATE\_ABSOLUTE\_COUNTER} to define a custom counter and \textit{ANNOTATE\_COUNTER\_VALUE}. to associate a power value to this counter.  
\item We use the command-line interface of ARM Streamline to begin a capture session. A capture session begins the process of data collection remotely from the target device. Before starting a capture session, the target device has to be set up with the gator daemon (Refer Section 3)  and the C++ executable to read the custom counters. On the host side, the configuration for a \textit{capture session} is defined in a $session.xml$ file which defines the target device IP address, the sampling frequency, resolution and the chosen hardware counters. The sampling frequency which can be set to one of three provided choices: Normal=1kHz, Low=100Hz and None. For experiments involving this tool, we set the sampling rate to Normal with resolution set to high to provide higher decimal precision of the power values collected. 
\end{itemize}
\subsection{Post processing step}
Python scripts were developed to process the measured power data into meaningful energy measurements. The ARM Streamline tool does not associate the power values to the code-annotations explicitly. Therefore, we  matched the collected function markers to the corresponding power values using time stamps.

For baseline energy of the GPU, the instantaneous power values were captured during the 10s sleep call. The total baseline energy for the total time ($T=10s$), was then calculated using Equation \ref{eq1} given in Section 3. Dividing total baseline energy by total time, the total baseline power for a single run was calculated. This was repeated for 10 runs and the average baseline power was considered ($0.06 \pm 0.02W$).
Similarly, for all other energy measurements, the energy was calculated using Equation \ref{eq1} using the desired time interval for either the entire application, a specific phase or a specific layer and the energy was reported for the minimum execution time across the $n$ runs. This choice is made based on the observation that the run with minimum execution time does not always correspond to minimum energy.  Error bars were plotted based on the variation from minimum to maximum energy measured across runs. Variation in the distribution of values exists across different sets of runs and we report the energy in the context of this variation.
\section{Performance and Energy Measurement}
The current methodology is designed to exploit the on-board power sensors to obtain energy measurements and this section is intended to demonstrate its usefulness and limitations to evaluate existing ConvNet algorithms on the Jetson TX1 and Caffe software framework. Such an evaluation could be done for other platforms such as Snapdragon \cite{qualnpe}. However, here we restrict ourselves to evaluation within this context.
\subsection{Component level Measurement versus system level measurements}
\begin{figure}[!t]
\centering
\includegraphics[width=\columnwidth]{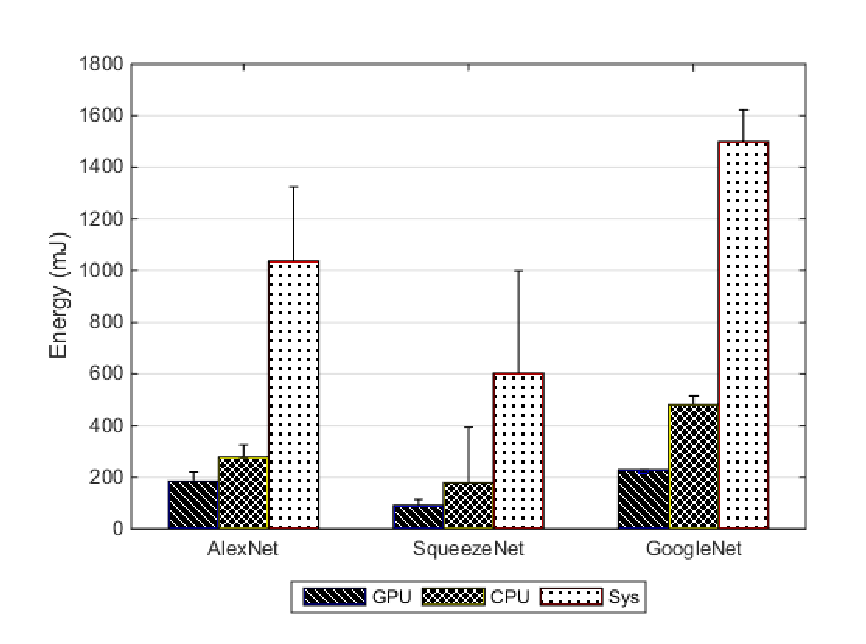}
\caption{Energy for inference step on GPU, measured at CPU, GPU and System level}
\label{fig:comp}
\end{figure}

As an example, to study the energy consumption at a component level, we sample the post-regulation power to the GPU. Figure \ref{fig:comp}, shows the performance and energy on our system at different levels (CPU, GPU and System) for an inference executing on the GPU. Here, the measurements are extracted for the inference phase of the application without any set up time in Caffe.  Since the baseline power is very small, see Subsection 4.3, we consider the energy measurement without baseline. We compare the execution behaviour of three popular Convnet models (AlexNet, SqueezeNet and GoogleNet) whose relative energy consumption have been compared previously \cite{yangdesign}. Here, the authors build a theoretical model of energy consumption based on the number of MACs and memory accesses for a specialized hardware platform to predict the energy-use of these ConvNets. However, such a energy estimation model does not exist for our system and relies on experimentation to understand if such models exhibit similar energy consumption characteristics. On our system, we obtain an energy consumption of 3.4x for ResNet50 compared to GoogleNet, 1.1x for GoogleNet compared to AlexNet. On comparison with results in \cite{yangdesign}, these models exhibit similar behaviour to the reported results. However, our results show inconsistencies with energy \textit{estimates} for SqueezeNet. The authors of \cite{yangdesign} report a higher energy consumption for SqueezeNet compared to AlexNet, whereas we report 1.9x for AlexNet compared to SqueezeNet.

In the given scenario, our power measurements were obtained only for the GPU which does not include the energy due to memory accesses. This may skew the interpretation of a result in favour of a given system.  Studies focused on comparing execution behaviour on different system \cite{eie}, for example, an inference on the CPU versus that on the GPU, should provide an upper bound to the energy consumption on the GPU by including related CPU energy consumption as well. This is because on our system, even if the application executes on the GPU, the CPU is still required to drive the GPU during that duration. Since, ConvNet models have storage costs (refer Section 2), including the associated energy due to memory is an important factor. However, on some devices it may not always be possible to measure component-level power at the CPU only or GPU only or DRAM only. In such scenarios, the highest level of power that can be measured would be the System-level power. For the power monitor on the TX1, we observe that the System power may equal the sum of the component power measurements plus the energy consumed by other system resources such as memory.  
This suggests that \textit{energy measurement} has several methodological choices to be made and is a time-consuming process. However, a systematic methodology guarantees that the actual execution behaviour on the given platform will be captured. 
 \textit{Energy estimation models}, on the other hand, relies on accurately capturing the execution behaviour of the application on the platform and may not always be applicable for all systems. In our work, we explore the possibility of developing an energy-prediction model for ConvNets based on \textit{actual} energy measurements on the Jetson TX1 as an alternative. We use a set of ConvNet models to create an energy predictor that can then be used to predict the energy consumption of other unseen ConvNets.
\subsection{Coarse-grained versus fine-grained application measurements}
\begin{figure}[!t]
\centering
\includegraphics[height=5.0cm]{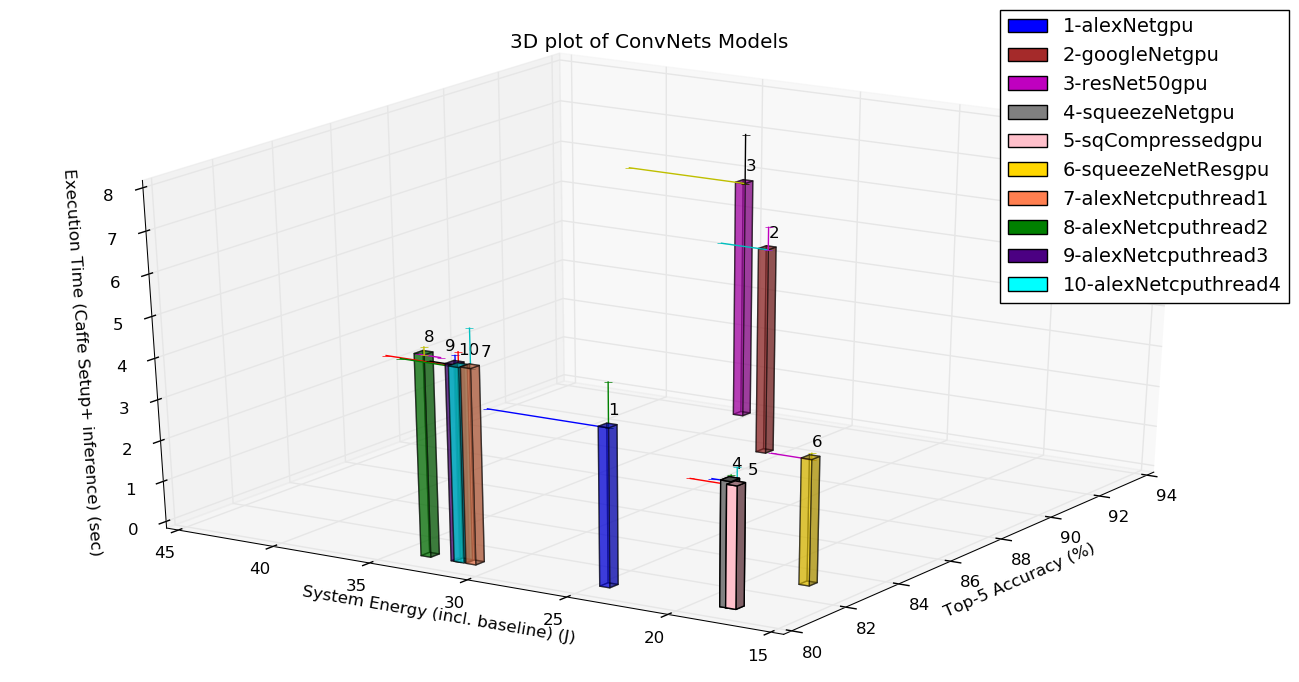}
\caption{System total energy + Baseline for inference step on CPU and GPU}
\label{fig:sysenergy}
\end{figure}

Figure \ref{fig:sysenergy} shows a \textit{coarse-grained level} system energy and execution time of the overall ConvNet inference. This includes Caffe's set up for model initialization and the inference phase. The top-5 accuracy is reported as published in the literature. Here, the system level energy is composed of the power consumed by the CPU, GPU and other system resources such as memory and any disk access. 
To illustrate our results, we  plot each evaluated ConvNet model in a 3D bar plot visualization with actual energy, time and accuracy measurements. Each bar represents a model with a colour and number code. Error bars\footnote{Error bars indicate maximum variation from minimum execution time} are plotted along the time and energy axes. Such a visualization puts into perspective the relative performance of each ConvNet on the metrics of time to solution, energy and accuracy on the given platform. For example, keeping in mind the set of models chosen for this study, we can find volumes of the graphs that are not populated such as the high accuracy, low energy and low time regime. If we systematically make trade-offs for a given model along any axis dimension, it would correspond to a model moving in a volume of space around its current position. This figure also shows the performance in time and energy for AlexNet, which is well studied in the literature, for two situations: inference running on the GPU (bar 1 in Figure \ref{fig:sysenergy}) and inference running on the CPU (bars 7,8,9 and 10 where each bar represents $n$ CPU threads and $n$ varies from 1 to 4). The CPU (with 4 threads) consumes 1.3x more energy and is 1.23x slower in time when compared to the GPU. To exploit parallelism on the CPU, the computation in certain layers such as the conv and fc layers are commonly re-structured into matrix-matrix and matrix-vector operations and relies heavily on BLAS libraries (Such as ATLAS, OpenBLAS) to make effective use of the CPU. However, we observe that the performance and energy does not scale with the number of threads, owing to the small amount of computation in a single image inference.

We also compare different ConvNets executed on the GPU. For example, we observe similar performance of SqueezeNet model and a modified SqueezeNet (sqCompressed) with deep compression. In the latter, the inference is performed with the decompressed version of the model where the benefits of compression are only exploited before actual execution. Therefore, there is no clear performance benefits during execution. SqueezeNetRes in another variation of the original SqueezeNet model with modified residual connections \cite{squeezenet, resnet}. This optimizations gives SqueezeNet a higher accuracy with similar execution behaviour as the original model. We can thus evaluate which optimization techniques lead to better trade-offs in performance, energy and accuracy simultaneously.
\begin{table}[]
\centering
\caption{Correlation between measured time and energy}
\label{correlation}
\begin{tabular}{|l|l|}
\hline
\textbf{} & \textbf{\begin{tabular}[c]{@{}l@{}} Pearson's Correlation\\ coefficient\end{tabular}} \\ \hline
\textit{alexNetGPU} & 0.99 \\ \hline
\textit{googlenetGPU} & 0.80 \\ \hline
\textit{squeezeNetGPU} & 0.51 \\ \hline
\textit{Googlenetbatch16GPU} & 0.91 \\ \hline
\textit{googlenet1batch1CPU} & 0.99 \\ \hline
\end{tabular}
\end{table}

\begin{figure}[!t]
\centering
\subfloat[AlexNet]{\includegraphics[width=\columnwidth]{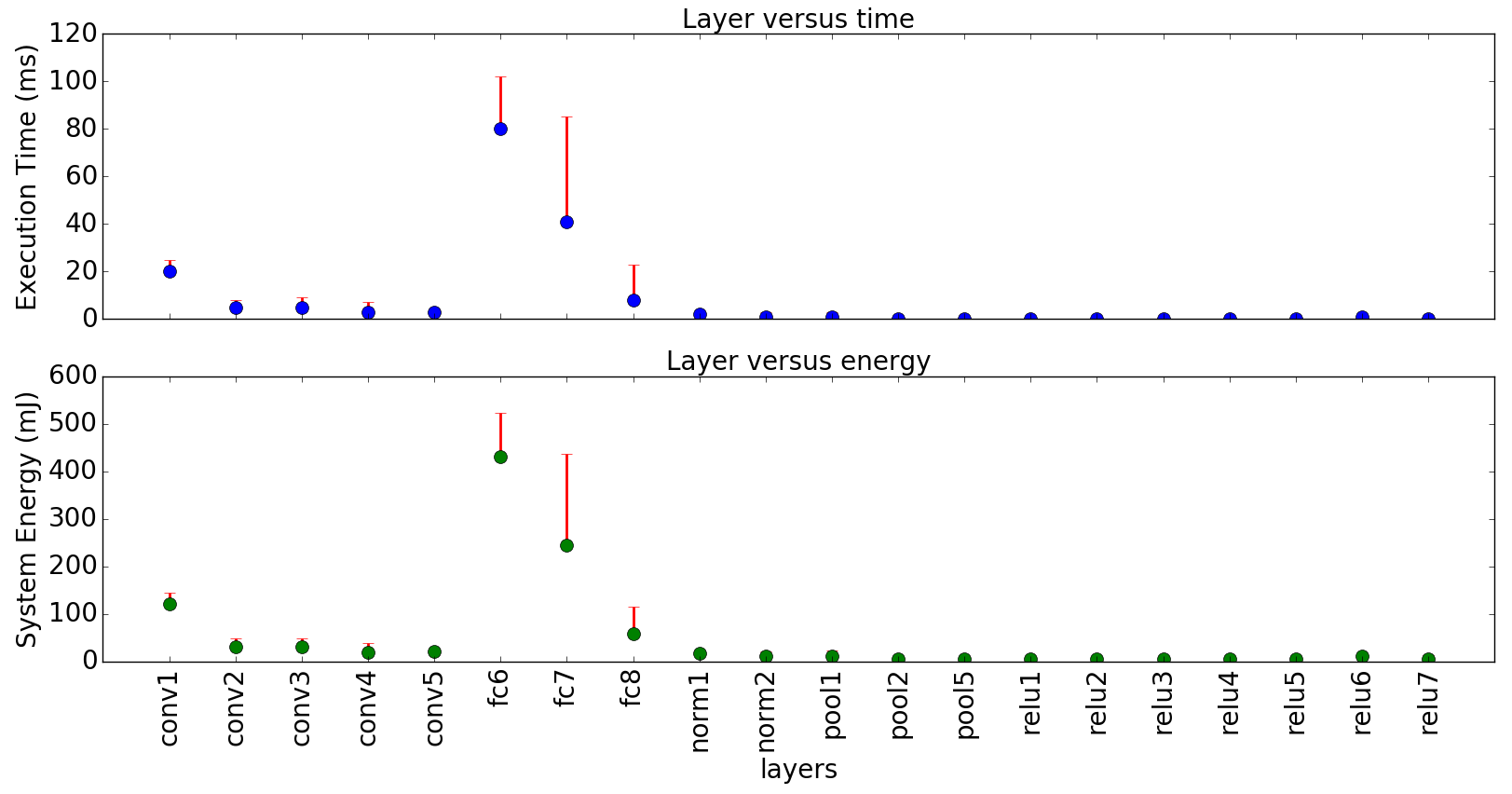}
\label{fig:layertrend1}}
\hfil
\subfloat[SqueezeNet]{\includegraphics[width=\columnwidth]{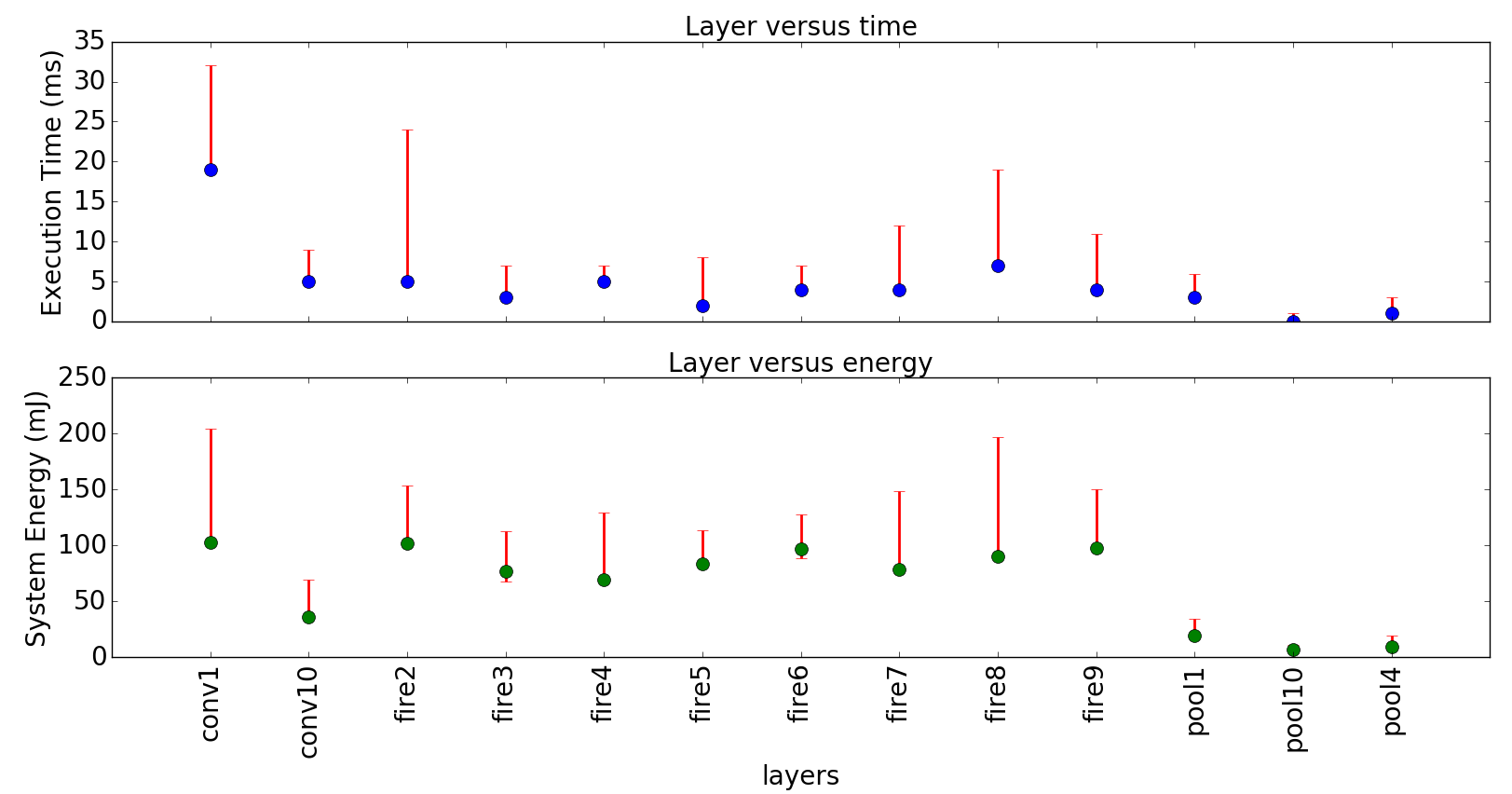}
\label{fig:layertrend2}}
\hfil
\subfloat[GoogleNet]{\includegraphics[width=\columnwidth]{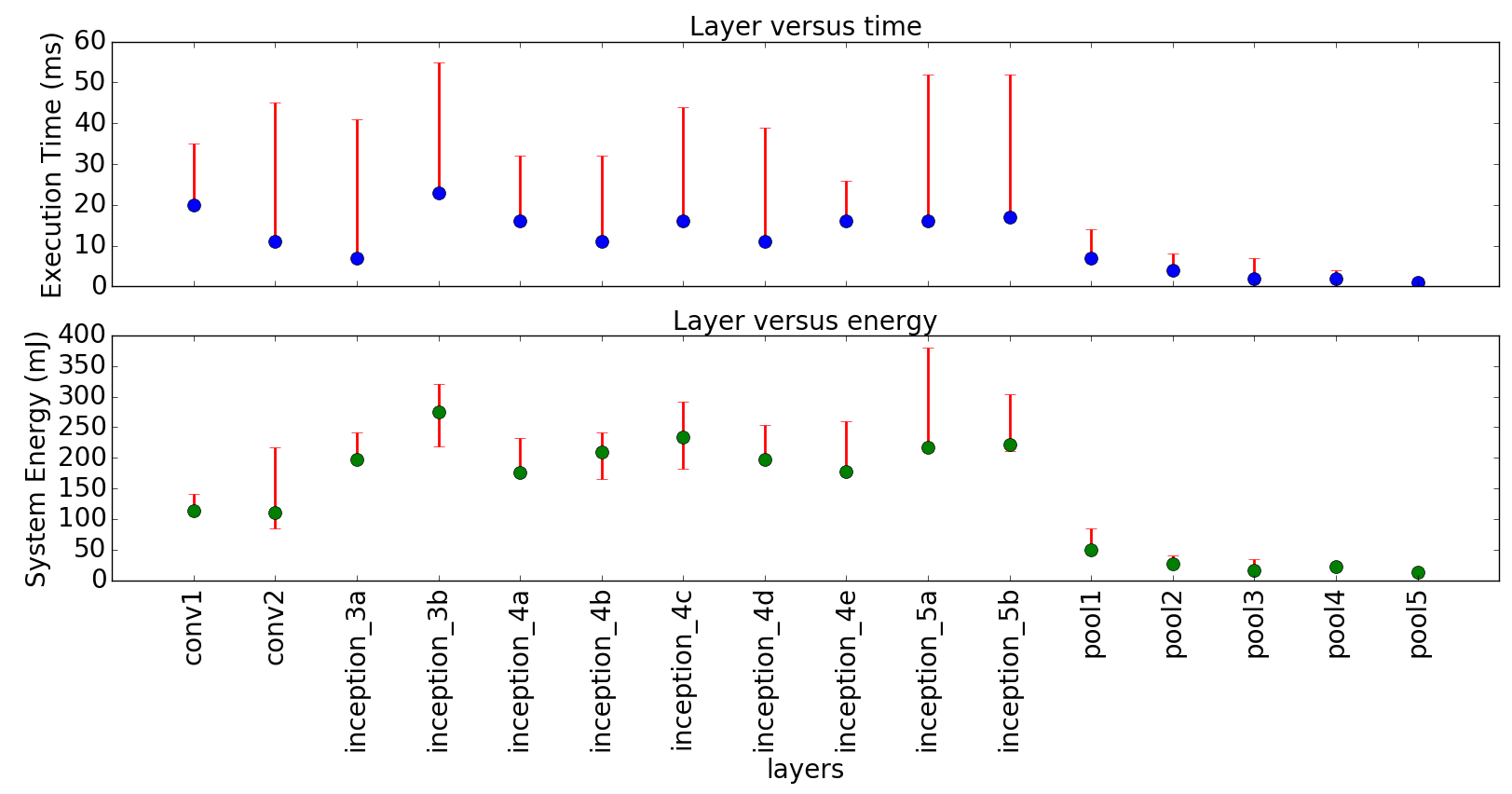}
\label{fig:layertrend3}}
\caption{Per layer performance and system energy profile of ConvNets with  batch 1 running on GPU}
\label{fig:layertrend}
\end{figure}

To study the execution behaviour at a \textit{fine-grained level} we extract per-layer system energy and performance measurements for each ConvNet. Figure \ref{fig:layertrend}, visualizes the per-layer execution behaviour of the selected ConvNet models (AlexNet, SqueezeNet and GoogleNet) for inferences executing on the GPU. Here, each point represents a top-level ConvNet layer type. The ordering of each layer along the x-axis is not representative of its actual execution order.  AlexNet was built using conv, pool, relu, norm and fc layers,  SqueezeNet encapsulates inside its \textit{fire} module a combination of $1\times1$ and $3\times3$ conv, each followed by relu layers and finally, GoogleNet has an arrangement of $1\times1$ and $3\times3$ conv filters along with max pooling into an inception module.
\begin{figure}
\centering
\includegraphics[width=\columnwidth]{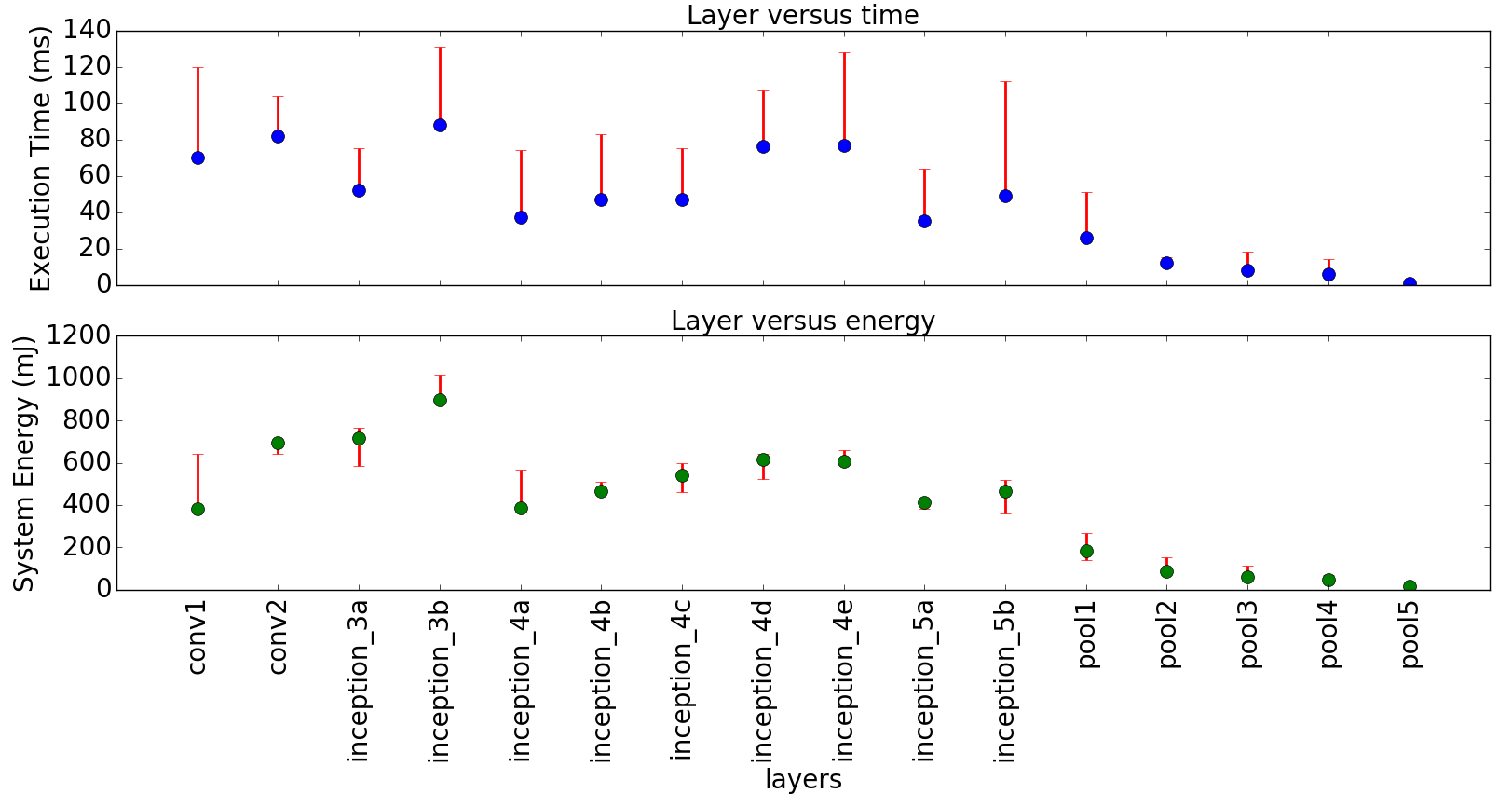}
\caption{Per layer performance and system energy profile for GoogleNet batch 16 inference running on GPU}
\label{fig:googlenet16gpu}
\end{figure}
From Figure \ref{fig:layertrend1} (AlexNet), we observe that time and energy are highly correlated with a Pearson's correlation factor of 0.99, as given in Table \ref{correlation}. Here, a fc layer that takes longer to execute and also consumes more energy. The fc layer relies on the matrix-vector implementation from the $CuBLAS$ library that may not have been optimized to account for the fact that mobile GPUs have limited GPU memory. Therefore, unlike performance benchmarks reported for desktop GPUs \cite{convnetbench,cnngpuperformance}, the smaller GPU memory results in spilling to main memory leading to high energy consumption and lower performance.

\begin{figure}
\centering
\includegraphics[width=\columnwidth]{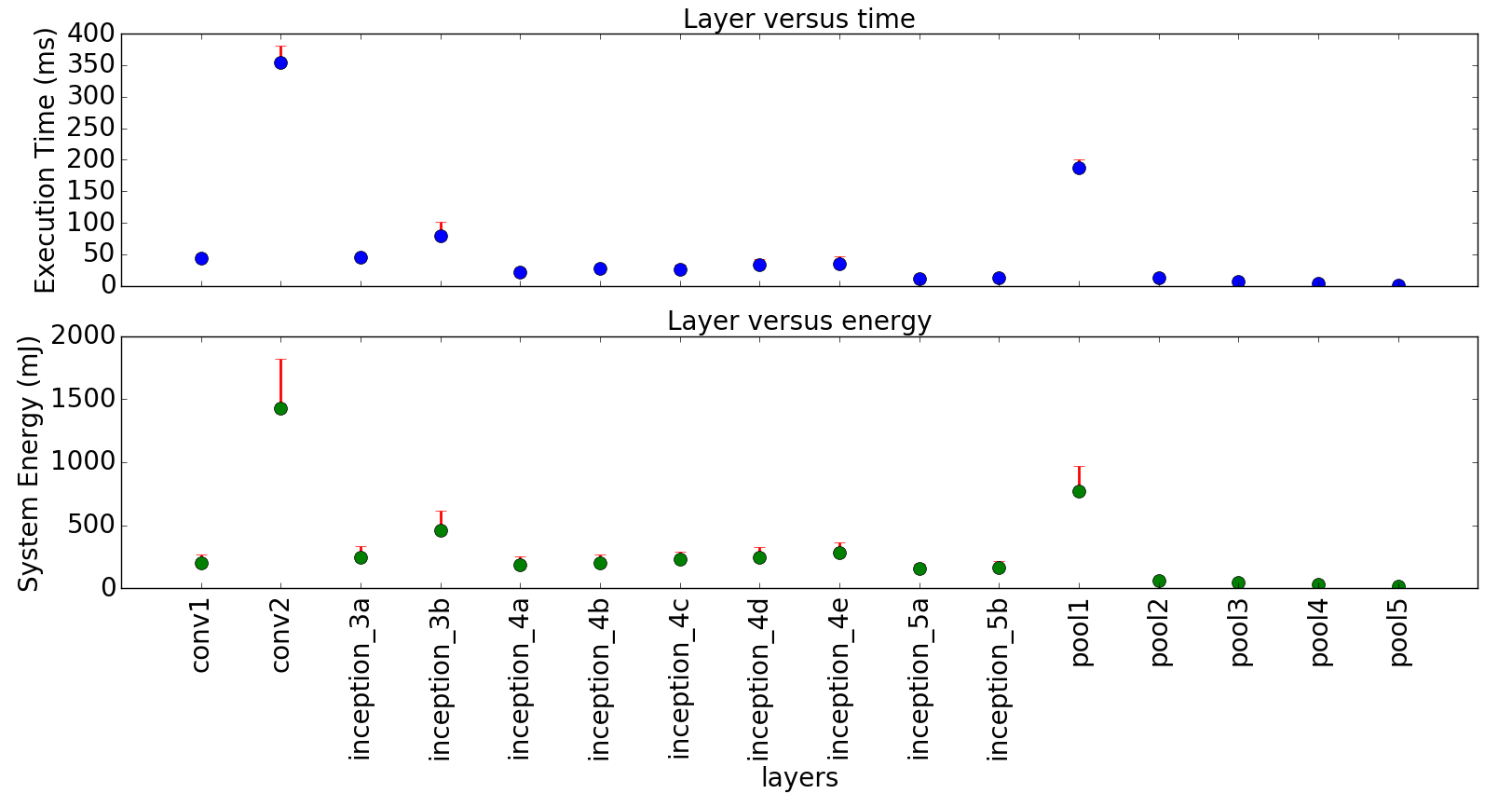}
\caption{Per layer performance and system energy profile for GoogleNet batch 1 inference running on CPU}
\label{fig:googlenetcpu}
\end{figure}

The correlation of time and energy start to diminish for models like GoogleNet and SqueezeNet on single image inferences, as seen in Figure \ref{fig:layertrend}. Taking GoogleNet as an example, we execute it with a larger batch size of 16. As GPUs are known to be very efficient for larger batch sizes \cite{nvidiawhite}, we observed in Figure \ref{fig:layertrend3} that the correlation in both time and energy starts to hold true again.
Finally, Figure \ref{fig:googlenetcpu} shows per-layer trends for GoogleNet with a batch size 1 running on the CPU. Here again, there is a strong correlation in time and energy indicating time could also be used as a proxy for energy. 
\section{Multi-Variable Regression model to predict energy consumption}
Given our ability to extract per-layer energy measurements, we demonstrate that it is possible to build an \textit{energy prediction} model for Convnets on the Jetson TX1 platform and Caffe framework. Our prediction model is based on the CPU measurements targeting its \textit{SIMD instruction} and \textit{bus accesses} performance counters.
This model is based on multi-variable linear regression, as given in \autoref{regression}, where the two dependent variables are SIMD instruction counts and number of bus accesses and the independent variable is \textit{Energy}. The Jetson TX1 system consists of a memory hierarchy with L1, L2 caches and main memory. However, for an initial prediction model, we restrict the prediction model to bus accesses (equivalent to last-level cache misses) as these are more expensive, in terms of energy, than cache accesses \cite{hanlearning}.
\begin{equation}
\label{regression}
\hat{E} = x_{1} \times  bus\_accesses_{conv} + x_{2} \times SIMD_{conv}
\end{equation}

The performance counters for SIMD instructions and bus accesses can be sampled using ARM's Streamline tool with code annotations to acquire per-layer counts. In our study, we focus on only the Conv layers for two reasons: The main component of a ConvNet is its Conv layers where current models, as given in \autoref{model}, tend to have multiple layers of Conv stacked on top of each other. This gives sufficient data points to build a sophisticated prediction model. Current research focuses on optimizing the Conv layers as they occupy 95\% of the execution time \cite{cnngpuperformance} which implies having a model to predict the cost of adding a Conv layer to the overall performance and energy consumption of a ConvNet should prove useful. The assumption here is that in our system, the Conv layers leverage standard optimized OpenBLAS routines to perform Matrix-Matrix multiplications on the CPU and CuDNN library for the GPU. For our CPU-based energy prediction model, we restrict the inferences to single-threaded executions on the CPU. One could think of more sophisticated models based on multi-threaded executions or a GPU. Similar experiments could be carried out for the GPU where additional executions would be required from tools such as \textit{nvprof} to obtain GPU-specific performance counters and correlating these to the collected energy values from separate runs.
\subsection{Energy prediction model from measured SIMD and bus accesses}
\label{mv}
We adopt the standard \textit{supervised learning} approach in machine learning to build a regression model \cite{mlprob}. In the training phase, the training data helps to establish a relationship, if any, between energy consumption and the two chosen performance counters. Once this prediction model is obtained, it can be tested on example ConvNet models not seen during the training phase.

Initially, the models from Table \ref{model} were qualitatively selected to form a training set on the following basis: AlexNet and VggNet-small represent structurally similar models that have fewer layers but relatively large number of parameters (62M-102M), SqueezeNet and SqueezeNetRes are an important class of ConvNets that are trained to keep the model size low without compromising accuracy (1.2M parameters, $\sim{80}$\% top-5 accuracy). GoogleNet has the best trade-off of size and accuracy (6.9M, 90.85\%) and ResNet50 represent the current state-of-the-art model in terms of accuracy in image classification tasks (93.2\%, 25M). SqueezeNet model with deep compression was excluded from the training set, as this is equivalent to a decompressed SqueezeNet model in terms of performance characteristics (refer Subsection 5.1).

As a first experiment, we focused on building and evaluating the robustness of the regression coefficients by training several prediction models on this set of ConvNets. The set of prediction models were obtained  by excluding a single ConvNet during the training phase. This is also known as \textit{Leave one out cross validation} \cite{mlprob}. Table \ref{tablecompare1} shows the prediction coefficients $x_1$ and $x_2$ derived for each excluded model. Here, each model is given in Column 1. Column 2, represents the coefficient for the total bus accesses ($x_1$) and Column 3 represents the coefficient for the total SIMD counts ($x_2$) in all the Conv layers for a given ConvNet. The first row represents a single experiment to form a regression model by excluding AlexNet as a data point during training. We then used the coefficients derived to obtain a training error which evaluates the regression model on the training set, as well a test error on the test set (data points excluded during training). We use the \textit{relative error} to quantify the performance of the predictor in relation to actual raw measurements obtained which is given by Equation \ref{relerr}: 
\begin{equation}
\label{relerr}
    Rel.Err (\%) = \frac{(|predicted\_value - actual\_value|)}{actual\_value} \times 100
\end{equation}

\begin{figure}
\centering
\includegraphics[width=\columnwidth]{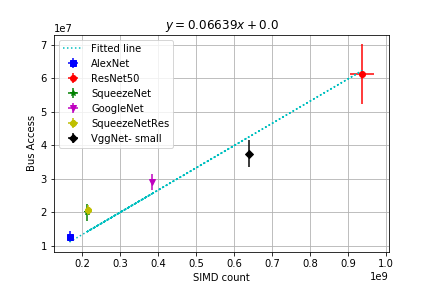}
\caption{Bus Access versus SIMD count}
\label{busSIMD}
\end{figure}

For example, in the first row, the average relative training error for the all ConvNets (which does not include AlexNet) in the training set is 5.36 $\pm$ 3.36\% and its test error on AlexNet is $2.23$\%. The average represents the bus access and SIMD counts over 5 separate runs and is shown in Figure \ref{busSIMD}. The corresponding average measured energy and average measured time is given in Column 5 and Column 6 of Table \ref{tablecompare1}. We provide timing measurement here for the sake of completeness. The value of each coefficient represents how strongly dependent the independent coefficient is on each of its dependent variables. We find that the coefficient for bus accesses contributes greater to the energy consumption which is consistent with the fact that accessing main memory cost more energy than the execution of a SIMD operation \cite{hanlearning}. 

For most cases, the average relative training error (that is, $4.81 \pm 3.19$\%) by including all the ConvNets (or allNets) in the training set is lower than by excluding individual models. \textit{Given the scenario where we can measure performance counters such as SIMD instruction and bus accesses, we are able to predict the energy consumption of unseen test ConvNets with an average relative test error of approximately $\sim{8\%}$ compared to actual energy measurements}.

Finally, to alleviate this need of having to measure SIMD and bus accesses counts, we attempted to explore the possibility of building prediction models for the two dependent variables $bus\_accesses_{conv}$ and $SIMD_{conv}$ themselves. This data was then fed into our current prediction model based on \textit{allNets} to obtain a final \textit{estimate} of energy consumption of any given ConvNet on the CPU of the Jetson TX1 platform, as discussed in the next subsections.
\begin{table*}[]
\centering
\caption{Regression Model to predict Energy}
\label{tablecompare1}
\resizebox{0.9\textwidth}{!}{%
\begin{tabular}{|l|l|l|l|l|l|l|l|}
\hline
 & \textbf{Bus access (x1)} & \textbf{SIMD (x2)} & \textbf{\begin{tabular}[c]{@{}l@{}} Predicted \\ Energy (mJ) ($\hat{E}$)\end{tabular}} & \textbf{\begin{tabular}[c]{@{}l@{}}Avg.  Measured \\ Energy (E) (mJ)\end{tabular}} & \textbf{Measured Time (sec)} & \textbf{\begin{tabular}[c]{@{}l@{}}Avg. Relative \\ Train error (\%)\end{tabular}} & \textbf{\begin{tabular}[c]{@{}l@{}}Relative \\ Test error (\%)\end{tabular}} \\ \hline
\textit{alexNet} & 3.37E-05 & 3.16E-06 & 951.28 & $ 930.44 $ & 0.1682 & 5.36 $\pm$ 3.36 & 2.23 \\ \hline
\textit{resNet50} & 3.89E-05 & 2.47E-06 & 4686.75 & $5261.42 $& 0.9468 & 2.03$\pm$ 2.06 & 10.92 \\ \hline
\textit{squeezeNet} & 4.09E-05 & 2.70E-06 & 1388.74 & $1240.29 $ & 0.2652 & 5.26 $\pm$ 1.88 & 11.96 \\ \hline
\textit{googleNet} & 3.76E-05 & 2.93E-06 & 2212.37 & $2072.48$ & 0.4228 & 5.76 $\pm$ 3.58 & 6.74 \\ \hline
\textit{squeezeNetRes} & 3.30E-05 & 3.20E-06 & 1365.02 & $1371.62 $& 0.2558 & 5.66 $\pm$ 2.5 & 0.48 \\ \hline
\textit{vggNet-small} & 1.27E-05 & 4.75E-06 & 3509.11 & $3027.99 $ & 0.5646 & 3.41 $\pm$ 2.67 & 15.88 \\ \hline
\textit{allNets} & 3.34E-05 & 3.18E-06 & \multicolumn{3}{l|}{} & 4.81 $\pm$ 3.19 &  \\ \hline
\textbf{\begin{tabular}[c]{@{}l@{}} Avg. Rel. Test Error  \\ (excluding a ConvNet)\end{tabular}} & \multicolumn{6}{l|}{\textbf{}} & \textbf{8.04 $\pm$ 5.96} \\ \hline
\end{tabular}
}
\end{table*}

\subsection{Predicting Conv layer SIMD counts}
\begin{figure}
\centering
\includegraphics[width=\columnwidth]{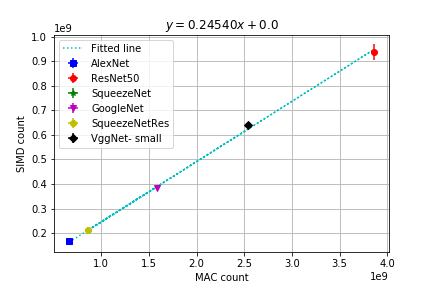}
\caption{SIMD versus MAC counts}
\label{macSIMD}
\end{figure}

SIMD operations exploit the data parallelism in Matrix-Matrix multiplication to obtain higher efficiency. Since the computation in a Conv layer can be transformed to a Matrix-Matrix multiplication operation, we explore the relationship between the application MAC count and its measured SIMD instructions for every Conv layer.
Given a description of the configurations of every layer in the Caffe's model prototxt file (See Section 4.1), we use Equation \ref{compeq1} in Section 2, to determine the MAC count for a Conv layer. Note, that for certain ConvNets AlexNet and MobileNet Equation \ref{compeq3} was used instead. The total MAC for all the Conv layers in a given ConvNet is tabulated in Column 3 of Table \ref{macSIMD}. This MAC count for all the Conv layers can then be used to build a simple linear regression model, as given in Equation \ref{simdEq}, to predict the SIMD counts for all the Conv layers in a ConvNet. The data corresponding to the measured SIMD ($y$) and predicted SIMD ($\hat{y}$) counts are shown in Column 2 and 4 of Table \ref{tablecompare2}. We use the same training set from subsection \ref{mv}, to build a simple SIMD predictor from MAC counts.
\begin{equation}
\label{simdEq}
    \hat{y}= c_{1} \times x
\end{equation}
We obtain a slope of 0.24, as show in Figure \ref{macSIMD}, which confirms that the SIMD width is 4 for the ARM CPUs on the Jetson TX1.\textit{Therefore, given an appropriate calculation of MAC count from the application, we can build a SIMD predictor that obtains an average relative test error of $0.65 \pm 0.94$\% compared to actual SIMD measurements}. Considering all the ConvNets the average relative test error is $1.06 \pm 0.80$\%
\begin{table}[]
\centering
\caption{SIMD prediction table}
\label{tablecompare2}
\resizebox{0.5\textwidth}{!}{%
\begin{tabular}{|l|l|l|l|l|}
\hline
                                       & \textbf{\begin{tabular}[c]{@{}l@{}}Avg. Measured (y) \\ SIMD \end{tabular}} & \textbf{MAC (x)} & \textbf{Predicted SIMD ($\hat{y}$)} & \textbf{\begin{tabular}[c]{@{}l@{}} Relative\\  error (\%)\end{tabular}} \\ \hline
\textit{alexNet}                       & 166326858                                                              & 665784864     & 163383605               & 1.76                                                                    \\ \hline
\textit{resNet-50}                     & 936965249                                                              & 3855925248    & 946244055               & 0.99                                                                    \\ \hline
\textit{squeezeNet}                    & 212510630                                                              & 861339936     & 211372820               & 0.53                                                                    \\ \hline
\textit{googleNet}                     & 383528521                                                              & 1581647872    & 388136387               & 1.20                                                                    \\ \hline
\textit{squeezenetRes}                 & 213932097                                                              & 861339936     & 211372820               & 1.19                                                                    \\ \hline
\textit{vgg -small}                    & 638627941                                                              & 2541337632    & 623644254               & 2.34                                                                    \\ \hline
\multicolumn{5}{|l|}{\textbf{Test Set}}                                                                                                                                                                                             \\ \hline
\textit{MobileNet-224}                 & 139589662                                                              & 567716352     & 139317592               & 0.12                                                                    \\ \hline
\textit{Places-CNDS-8s}           & 492978185                                                           & 1967702016    & 482874074               & 2.04                                                                    \\ \hline
\textit{ALL-CNN-C}                     & 66909070                                                             & 270798336     & 66453911                & 0.37                                                                    \\ \hline
\textit{Inception-BN}                  & 834842927                                                            & 3400527872    & 834489539               & 0.02                                                                    \\ \hline
\textbf{Avg. Relative Test Error (\%)} & \multicolumn{3}{l|}{\textbf{}}                                                                                   & \textbf{0.65 $\pm$ 0.94}                                                   \\ \hline
\end{tabular}
}
\end{table}

\subsection{Predicting Conv layer bus accesses}
Conv layers are often preceded and succeeded by data transformation operations such as \textit{im2col}  to transform it into a Matrix-Matrix computation, and \textit{col2im} to transform it back into the original 2D image layout \cite{cudnn}. This is because these Conv layers are interleaved with pooling, relu and other layers which required data in the specific 2D format. Each ConvNet model differs in terms of how it is interleaved (See Section 2 on Conv layer topologies). Even though, we can calculate the bandwidth of each Conv layer as given in Equation \ref{compeq2}, the data re-structuring and associated complex cache memory hierarchy make the relationship between data movement between layers and performance counters such as cache and bus accesses non-trivial. 

However, somewhat surprisingly we found that a linear relationship exists between the total number of measured bus accesses and SIMD counts in the Conv layers, which can be seen in Figure \ref{busSIMD}. Therefore, a similar linear regression predictor, as given in Equation \ref{simdEq}, was built to determine the bus access counts for the Conv layers from the measured SIMD counts. We find a linear relationship exists with a line of slope of 0.0663 between SIMD and bus access counts. We then use the predicted SIMD ($\hat{y}$) counts obtained previously to predict the bus access counts ($\hat{z}$) for all the ConvNets, as given in Column 4 of Table \ref{tablecompare3}. \textit{ For most ConvNets we are able to obtain a good prediction of bus access counts from SIMD with an average relative test error of $17.09 \pm 13$ \%}. However, three of the ConvNets from the original test set exhibit a spike in individual relative errors with MobileNet around $\sim 73 \%$ while the other two are below 50\% ( All-CNN-C around $\sim 38\%$ and squeezeNetRes $\sim 32\%$). Our predicted results for bus accesses from predicted SIMD counts by including MobileNet increases the avearage relative test error by $1.3x$ when compared to excluding MobileNet. We hypothesize that MobileNet exhibits different-to-typical characteristics in its data access patterns that cannot be trivially predicted from SIMD counts alone and is left for future exploration.

\begin{table}[]
\centering
\caption{Bus Access prediction table}
\label{tablecompare3}
\resizebox{0.5\textwidth}{!}{%
\begin{tabular}{|l|l|l|l|l|}
\hline
 & \textbf{\begin{tabular}[c]{@{}l@{}}Predicted \\ SIMD ($\hat{y}$) \end{tabular}} & \textbf{\begin{tabular}[c]{@{}l@{}}Avg. Measured \\ Bus access (z) \end{tabular}} & \textbf{\begin{tabular}[c]{@{}l@{}}Predicted Bus\\ Access ($\hat{z}$) \end{tabular}} & \textbf{\begin{tabular}[c]{@{}l@{}}Relative\\  error (\%)\end{tabular}} \\ \hline
\textit{alexNet} & 166326858 & 12635625 & 10847037 & 14.15 \\ \hline
\textit{resNet-50} & 936965249 & 61100440 & 62821142 & 2.81 \\ \hline
\textit{squeezeNet} & 212510630 & 19929941 & 14033041 & 29.58 \\ \hline
\textit{googleNet} & 383528521 & 28927569 & 25768374 & 10.92 \\ \hline
\textit{squeezenetRes} & 213932097 & 20600111 & 140330412 & 31.87 \\ \hline
\textit{vgg - small} & 638627941 & 37448187 & 41403742 & 10.56 \\ \hline
\textit{MobileNet-224} & 139589662 & 34642804 & 9249294 & 73.30 \\ \hline
\textit{Places-CNDS-8s} & 492978185 & 31498902 & 32058009 & 1.77 \\ \hline
\textit{ALL-CNN-C} & 66909070 & 7172165 & 4411875 & 38.48 \\ \hline
\textit{Inception-BN} & 834842927 & 64169256 & 55401760 & 13.66 \\ \hline
\textbf{\begin{tabular}[c]{@{}l@{}}Avg. Rel. Test \\ Error w/o MobileNet (\%)\end{tabular}} & \multicolumn{3}{l|}{\textbf{}} & \textbf{17.09 $\pm$ 13} \\ \hline
\textbf{\begin{tabular}[c]{@{}l@{}}Avg. Rel. Test Error \\ with MobileNet (\%)\end{tabular}} & \multicolumn{3}{l|}{} & \textbf{22.71 $\pm$ 21.6} \\ \hline
\end{tabular}
}
\end{table}
\subsection{Energy prediction model from simulated data}
Our final step, is to estimate the energy consumption of all the Conv layers in a ConvNet directly from the application parameters, which in our case is the Conv layer MAC count.
The prediction for the SIMD ($\hat{y}$) and bus access counts ($\hat{z}$) for the Conv layers can now be fed into our initial energy prediction model, given in Subsection 6.1. We consider the regression coefficients derived from using \textit{allNets}. This data is tabulated in Table \ref{tablecompare4}, where the predicted energy ($\hat{E}$) is given in Column 2. \textit{Therefore, by excluding MobileNet, we are able to predict the energy consumption of the Conv layers of any given ConvNet, solely using MAC count, with an average relative error test rate of   $7.08\pm 6.0\%$.}

\begin{table*}[]
\centering
\caption{Energy Prediction Results for coeffs. $x_1 =3.34E-05$ and $x_{2} = 3.18E-06$ }
\label{tablecompare4}
\resizebox{0.8\textwidth}{!}{%
\begin{tabular}{|l|l|l|l|l|}
\hline
 & \textbf{\begin{tabular}[c]{@{}l@{}}Predicted \\ Energy ($\hat{E}$) (mJ)\end{tabular}} & \textbf{\begin{tabular}[c]{@{}l@{}}Average Measured \\ Energy (E) (mJ)\end{tabular}} & \textbf{\begin{tabular}[c]{@{}l@{}}Average Measured \\ Time (sec)\end{tabular}} & \textbf{\begin{tabular}[c]{@{}l@{}}Relative \\ error (\%)\end{tabular}} \\ \hline
\textit{alexNet} & 881.41 & 930.45 & 0.17 & 5.26 \\ \hline
\textit{resNet-50} & 5104.76 & 5261.42 & 0.95 & 2.97 \\ \hline
\textit{squeezeNet} & 1140.30 & 1240.30 & 0.27 & 8.06 \\ \hline
\textit{googleNet} & 2093.90 & 2072.49 & 0.42 & 1.03 \\ \hline
\textit{squeezeNetRes} & 1140.30 & 1371.62 & 0.25 & 16.86 \\ \hline
\textit{vgg - small} & 3364.41 & 3028.00 & 0.56 & 11.11 \\ \hline
\textit{Places-CNDS-8s} & 2604.99 & 2613.46 & 0.46 & 0.32 \\ \hline
\textit{ALL-CNN-C} & 358.50 & 422.29 & 0.08 & 15.10 \\ \hline
\textit{Inception-BN} & 4501.87 & 4641.14 & 0.84 & 3.00 \\ \hline
\textit{MobileNet} & 751.58 & 1824.60 & 0.35 & 58.80 \\ \hline
\textbf{\begin{tabular}[c]{@{}l@{}}Average Relative Test \\ Error (\%) w/o MobileNet\end{tabular}} & \multicolumn{3}{l|}{\textbf{}} & \textbf{7.08 $\pm$ 6.0} \\ \hline
\textbf{\begin{tabular}[c]{@{}l@{}}Average Relative Test \\ Error (\%) with MobileNet\end{tabular}} & \multicolumn{3}{l|}{\textbf{}} & \textbf{17.33 $\pm$ 12.2} \\ \hline
\end{tabular}
}
\end{table*}

\section{Related Work}
This section focuses on several benchmarking frameworks that exists for evaluation of deep learning models. We provide a snapshot view of the main purpose of each framework. Most benchmarks focus on either a single application domain on a specific bottleneck layer (for example, a conv layer) on a single platform  while others are generic, targeting either many application areas or different hardware systems. Moreover, most of them have relied solely on benchmarking of time rather than energy to characterize the execution behaviour of the application.

\textit{convnet-benchmark} \cite{convnetbench}, benchmarks all public open-source implementations of ConvNets. This provides overall and layer-wise timing benchmarks of the Convolution operation. However, this benchmarking work is carried for a single application area on a single desktop machine consisting of 6-core Intel Core i7 CPU and an NVIDIA Titan X GPU and does not include executions on embedded platform.

\textit{Fathom} \cite{fathom}, provides a comprehensive benchmarking suite to include state-of-the-art neural network models from both image processing and language processing domains. The authors focus on understanding the holistic execution behaviour in terms of execution time of these models while breaking down its execution behaviour in terms of high-level Tensorflow operations. All experiments were performed on the Skylake i7-6700k desktop CPU.

\textit{DAWNbench} \cite{dawn}, proposes an end-to-end performance evaluation of deep learning models. Rather than focusing on the performance of the computation within layers, the authors study the performance impact of the interaction between various optimizations techniques during training. For example, the impact on convergence rate during training with different batch sizes. Studies on the interaction of various energy-efficiency techniques such as low-precision, compression techniques and others, remain unexplored and we envision the possibility of such studies with the development of better energy benchmarking tools along the lines presented in this work.

Few studies have emerged that report energy and performance of deep learning models on the TX1 platform \cite{nvidiawhite, eie, analysis}. However, these studies are often adhoc to study a limited set of deep learning models and platform-to-platform comparisons. These studies lack in a consistent methodology to acquire power for the Jetson TX1 and provide minimum details of the adopted method. Our work instead develops a detailed methodology to acquire power measurements using the power sensor on-board the TX1.

Similar approaches to our energy profiling approach exist that provide energy consumption at a functional level \cite{hotspots,energyml}. In \cite{energyml}, the authors use code-annotations to demarcate specific phases of a decision-tree machine learning model. However, our work is aimed towards understanding fine-grained energy consumption of neural network algorithms.

BenchIP \cite{benchip} is an industry-level benchmark suite and methodology that evaluates the efficiency deep learning workloads comprising ConvNets and Long Short Term Memory networks (LSTM) models on representative platforms from desktop, server and embedded domains. The authors evaluate each layer in isolation as well as end-to-end model executions in Caffe. However, it differs from our approach, where we study each layer in isolation in the context of an actual inference. To the best of our knowledge, this benchmark suite is yet to be open-sourced.  

While benchmarking efforts are continuing to grow, it is a time-consuming effort. Therefore, energy estimation tools have been proposed to evaluate deep learning models \cite{yangdesign}.
This work provides a model of energy consumption based on the number of MACs and bandwidth, and associates a hardware cost of each to estimate the energy consumption of a neural net. However, this estimation methodology is developed for specialized dataflow implementations on a specialized hardware accelerator "Eyeriss" \cite{eye} and may not be representative for models executing on other platforms. Our work takes a similar approach by building a model of energy consumption using only MAC counts and regression analysis to build an initial energy estimation model for thev CPU on the Jetson TX1 platform.

\section{Conclusions and Future Work}
Deployment of deep learning applications on mobile and embedded platforms remains a challenge due to limited power budgets available on such devices. Efforts to improve energy consumption of deep learning applications have begun to emerge from the development of compact ConvNet models to building specialized hardware.
Existing benchmarking efforts tend to characterize the execution behaviour of deep learning applications on high-end desktop CPUs and GPUs and often neglect embedded platforms. Therefore, we present "SyNERGY", a framework to enable measurement of \textit{fine-grained} performance and energy of deep learning applications targeting embedded platforms such as the Jetson TX1. We demonstrate a systematic methodology using its power monitoring sensor (TI-INA3221x) and integrate Caffe and ARM's Streamline Performance Analyser, to enable coarse-grained and fine-grained energy profiling for single image inferences. We report energy measurements for several popular ConvNets such as AlexNet, SqueezeNet and GoogleNet for an entire inference, specific layers and at different levels such as CPU, GPU and System. 

Through our benchmarking framework, we were able to build an initial energy prediction model using 11 representative ConvNet models. Our initial energy prediction model was based on data gathered from SIMD and bus access CPU performance counters with actual execution runs of these models. We are able to predict the energy-use of all the Conv layers in a ConvNet model with an average relative test error rate of $8.04 \pm 5.96$\% over actual energy measurements.
Furthermore, we build upon this model to make predictions directly from the application parameters. Our predictor achieves $17.33 \pm 12.2$\% (or $7.08 \pm 6.0$\% if we exclude MobileNet) average relative test-error by using only Multiply-Accumulate (MAC) counts calculated directly from the application description as input to predictors for SIMD and bus accesses. 

Future work, includes extending this energy prediction model to other layers in the model such as fully-connected layers such that we can make energy predictions for an \textit{entire} ConvNet. This opens up the possibility of targeting other deep learning models such as those in natural language processing domains.
We also aim to build an energy predictor for deep learning applications on other embedded platforms such as the Snapdragon.

In terms of the quality of the energy-predictor itself, improvements to the predictor can be made by targeting other performance counters by including L1 and L2 cache access. Opportunities for in-depth analysis work to guide the use of power management techniques such as DVFS to reduce energy consumption at specific layers could also be explored.

\section{Acknowledgements}